\documentstyle[12pt]{article}

\begin{document}

\baselineskip = 18pt

\def\la{\mathrel{\mathpalette\fun <}}
\def\ga{\mathrel{\mathpalette\fun >}}
\def\fun#1#2{\lower3.6pt\vbox{\baselineskip0pt\lineskip.9pt
  \ialign{$\mathsurround=0pt#1\hfil##\hfil$\crcr#2\crcr\sim\crcr}}}

\hskip 4 in OSU-TA-17/95

\vskip 0.6 cm

\centerline{\bf Differential neutrino heating and reduced $^4$He production}

\centerline{\bf from decaying particles in the early universe}

\vskip .6 cm
\centerline{Richard E. Leonard and Robert J. Scherrer}
\centerline{\it Department of Physics, The Ohio State University,
Columbus,  Ohio 43210}
\vskip .6 cm

\centerline{\it Abstract}
{\narrower  A decaying particle that dominates the energy density of
the universe and then decays with a lifetime $\sim 10^{-1}$ sec into
electromagnetically-interacting
particles can heat the background electron
neutrinos more than the $\mu$ and $\tau$ neutrinos, resulting in a decrease
in the primordial $^4$He abundance.  Previous calculations of this effect
assumed sudden decoupling of the neutrinos from the
electromagnetic plasma.  Here
we repeat this calculation using an improved (although still approximate)
treatment of the energy flow from the blackbody radiation into the various
neutrino components.  We find that with our improved calculation,
the reduction in the primordial $^4$He is not as large as had been previously
thought, but it occurs over a larger range in decaying particle lifetime
$\tau$.
The reduction in $^4$He
can be as large as $\Delta Y = -0.01$ for $\tau = 0.1 - 0.7$ sec.

\vfill
\eject

\noindent {\bf I. INTRODUCTION}

\indent  The effects of decaying particles on primordial nucleosynthesis
have been investigated by many authors (see, for example, [1] - [2] and
references therein).  One interesting effect, first noted by Scherrer \& Turner
[3] can occur if the particle decays into products which interact
electromagnetically and go into thermal equilibrium with the cosmic
blackbody background.  If such decays occur well after the neutrinos
have dropped out of thermal equilibrium (at a temperature of a few MeV)
then the neutrinos do not share in the heating of the blackbody background.
For lifetimes much shorter than the neutrino decoupling temperature,
the neutrinos share fully in the heating of the blackbody background.
In the intermediate regime (lifetimes on the order of 0.1 sec), an interesting
effect can occur.  Because the electron neutrino couples more strongly to the
$e^-e^+$ pairs
than do the $\mu$ and $\tau$ neutrinos, it decouples at a slightly lower
temperature.
Thus, for lifetimes in this intermediate range, it is possible for the decaying
particle to heat the electron neutrinos more than the $\mu$ and $\tau$
neutrinos.
This results in a net decrease in the expansion rate relative to
the weak ($n \leftrightarrow p$) rates, giving a reduction in the $^4$He
abundance [3].
This effect can occur, for example, in gravitino decay [4].

This possibility of reducing the primordial $^4$He production is quite
interesting,
because recent calculations suggest that the standard model of
big bang nucleosynthesis can be made consistent with current observations
only if the ``true" primordial $^4$He abundance is larger than currently
believed,
or if primordial nucleosynthesis is modified to give a smaller production of
$^4$He [5].
However, the treatment of this effect in references [3]-[4] was quite crude:
the neutrinos
were assumed to decouple suddenly when their interaction rates dropped
below the expansion rate; entropy dumped into the system before decoupling
was shared equally by the neutrinos, photons, and $e^-e^+$ pairs,
while entropy released after decoupling went entirely into the photons
and $e^-e^+$ pairs.  Hence, we are motivated to do a more careful
investigation,
involving a more accurate treatment of the transfer of energy from the decaying
particle to the neutrinos.  We use the linearized energy-transfer equation
given by Rana and Mitra [6], based on earlier work by Herrera and Hacyan [7].
We do not calculate the full distortion in the spectrum, but instead treat each
species of
neutrino
as a black body with a single temperature.  Although this is not as
sophisticated
as the treatment given by Dodelson et al. [2], it represents a significant
improvement over other previous studies of this problem.
(Dodelson et al. examined only
decaying neutrinos, which do not have a large enough energy density to make
this an interesting effect).

In the next section, we present our calculation of the evolution of the
neutrino temperature in the presence of a decaying particle with a short
($10^{-2} -$ 2 sec) lifetime.
In Sec. III, we calculate the effect of the differential neutrino
heating on the primordial $^4$He abundance, and briefly summarize our
conclusions.
We find that the reduction in $^4$He production is smaller than had been
previously
thought, but it is still large enough to be interesting.

\vskip 0.4 cm

\noindent {\bf II. EVOLUTION OF THE NEUTRINO TEMPERATURE}

The calculation begins by
adding to the standard cosmological model an X particle which
decays exclusively into particles that thermalize rapidly compared with the
expansion
timescale.  This particle
adds a contribution to the total energy density driving the expansion
according to
the Friedmann equation
\begin{equation}
H \equiv \frac{\dot{R}}{R} = (\frac{8}{3} \pi G \rho)^{1/2}.
\end{equation}
where R is the expansion scale factor.
Here the total density $\rho = \rho_{std} + \rho_{X}$ with $\rho_{std}$
representing the energy
density in the standard model.  The X density $\rho_{X}$ evolves according to
the equation
\begin{equation}
\dot\rho_{X} = -\Gamma \rho_{X} - 3H\rho_{X},
\end{equation}
where $\Gamma$ is the decay rate, $\Gamma \equiv 1/\tau$,
and $\tau$ is the X particle lifetime.  This equation has the solution
\begin{equation}
\rho_{X} = \rho_{X0}(\frac{R}{R_0})^{-3}e^{-t/\tau}.
\end{equation}
We consider particle lifetimes in the range $10^{-2} {\rm sec} < \tau
< 2$ sec.  Following reference [3], we quantify the
the particle number density
by the parameter $r$, which gives the number density of the X particles
relative to photons before $e^-e^+$ annihilation.  The only quantity
which determines the energy density of the X particles
as a function of temperature is $r m_X$, where
$m_X$ is the particle mass.  In all of our calculations,
we take $r m_X = 10^4$ MeV.  For this choice of $rm_X$ and particle lifetimes
above,
the density of the X particles
dominates the expansion prior to decay.  More importantly, the entropy
produced by the decay dominates the previously-existing radiation
during nucleosynthesis.  When these two conditions are satisfied,
the evolution of the element abundances is independent of $rm_X$,
and the final results will be essentially
a function only of $\tau$ [3].  For the range of lifetimes considered here,
these conditions will be satisfied for $r m_X \ga 10^3$ MeV, so
this is the range over which our results will be applicable (See ref. [3]).
For comparison, a massive tau neutrino has a maximum
$r m_X$ of about 1 MeV (for $m_\nu \sim 3-5$ MeV), so we are examining
a very different region of parameter space from that discussed in ref. [2].

The decay of the X particle heats the electromagnetic plasma.  Let
$\rho_\gamma$ and $p_\gamma$ be
the energy density and pressure of the photons and all relativistic
particles in thermal equilibrium with the photons.
Then the evolution of $\rho_\gamma$ is given by
\begin{equation}
\label{rhogamma}
\dot\rho_{\gamma} = \Gamma \rho_{X} - 3H(\rho_\gamma + p_\gamma).
\end{equation}

In references [3]-[4], the authors used a sudden decoupling approximation
to deal with neutrino heating.  In this approximation
the neutrinos are assumed to have blackbody spectra with temperatures
$T_{\nu_i}$, $i = e,\mu, \tau$.  The temperature of an individual neutrino
species
is set equal to the photon temperature until it
decouples and thereafter the neutrino temperature decreases as $1/R$.
The decoupling condition used in references [3]-[4]
is
$\Gamma_{\nu_i}/H < 1$ where $\Gamma_{\nu_i} = A_{\nu_i} T^5$
($i = e,\mu,\tau$).  For consistency with our derivations below,
we re-express $\Gamma_{\nu_i}$ in the form
\begin{equation}
\label{Gammaold}
\Gamma_{\nu_i} = \alpha_{\nu_i} G_F^2 T^5,
\end{equation}
where
$G_F$ is the Fermi coupling constant, and
$\alpha_{\nu_i}$ is a dimensionless constant of order unity.  The values for
$A_{\nu_i}$ used
in references [3]-[4] correspond to $\alpha_{\nu_e} = 0.3$ and
$\alpha_{\nu_{\mu,\tau}} =
0.07$.
This is obviously a very crude
approximation and is considered here only for comparison with our new
calculations.

In this work we follow the treatment of Rana and Mitra [6] and
Herrera and Hacyan [7],
who treated the neutrinos as a perfect blackbody and calculated the
rate of energy transfer from the electron-positron pairs to the neutrinos
through scattering and annihilation.  Let $u_{\nu_i}$ be the total rate
of energy density transfer from the $e^-e^+$ pairs to a single
neutrino species $\nu_i$ ($i = e, \mu, \tau$) via
annihilations and scatterings.  Then equation (\ref{rhogamma}) is modified to
\begin{equation}
\dot\rho_{\gamma} = \Gamma \rho_{X} - \sum_{i=e,\mu,\tau}
{u_{\nu_i}} - 3H(\rho+p)_{\gamma}
\end{equation}
while the neutrino density evolves as
\begin{equation}
\label{rhonu}
\dot\rho_{\nu_i} = u_{\nu_i} - 4H\rho_{\nu_i},
\end{equation}
In the limit of
small temperature difference between the electrons and neutrinos,
$u_{\nu_i}$ is given by [6]
\begin{equation}
\label{uvalue}
u_{\nu_i} = 2 I_{e^- \rightarrow \nu_i} (T_{\gamma}-T_{\nu_i})
\end{equation}
where $I_{e^- \rightarrow \nu_i}$ is a function of $T_\gamma$ which
gives the rate of energy density transfer per unit temperature difference from
electrons into a particular neutrino species $\nu_i$, and the factor
of 2 comes from including both electrons and positrons.
In the limit of large temperatures ($T_\gamma \gg m_e$), $I$ has
the form
\begin{equation}
\label{Ivalue}
I_{e^- \rightarrow \nu_i} = \beta_{e^- \rightarrow \nu_i} G_F^2 T_\gamma^8,
\end{equation}
where
$\beta_{e^- \rightarrow \nu_i}$ is a dimensionless constant
of order unity.  [Equation (\ref{Ivalue}) comes from the fact that
$I_{e^- \rightarrow \nu_i}$ is approximately equal to the interaction rate
($\Gamma = n\langle\sigma v\rangle \sim G_F^2 T^5$) times the heat capacity
($\sim T^3$).]
Three distinct processes contribute to the net value of
$I_{e^- \rightarrow \nu_i}$:  $e^-e^+$
annihilation into $\nu_i \bar \nu_i$ pairs (along with the inverse
annihilation), $e^-$ scattering off of $\nu_i$, and $e^-$ scattering
off of $\bar \nu_i$.
The values for $\beta_{e^- \rightarrow \nu_i}$ can be derived from the
high-temperature limit of
the
results of Rana and Mitra [6] and
are listed in Table 1, where we have included both the total $\beta_{e^-
\rightarrow \nu_i}$
used in equation (\ref{Ivalue}), as well as the contribution to $\beta_{e^-
\rightarrow \nu_i}$ from the individual reactions.  It is obvious from the
numbers in Table 1
that $e^-e^+$ annihilation is the dominant energy transfer mechanism.

\begin{table}[h]
\caption{The coefficients for the energy transfer rates in equations
(\ref{Ivalue}) and (\ref{Ivalue2}), where $\nu_i$ denotes a $\mu$ or $\tau$
neutrino.
We have included the breakdown of $\beta_{e^- \rightarrow \nu_i}$ into specific
reactions as well as the total
$\beta_{e^- \rightarrow \nu_i}$ used in equations (\ref{Ivalue}) and
(\ref{Ivalue2}).}
\begin{center}
\begin{tabular}{|r|r|r|r|r|} \hline
Process & $\beta$ \\ \hline
$e^-\nu_e \rightarrow e^-\nu_e$ & 0.14 \\
$e^-\bar\nu_e \rightarrow e^-\bar\nu_e$ & 0.04 \\
$e^- e^+ \leftrightarrow \nu_e \bar \nu_e$ & 0.99 \\ \hline
$e^- \rightarrow \nu_e$   &  1.17 \\ \hline
$e^-\nu_i \rightarrow e^-\nu_i$ & 0.02 \\
$e^-\bar\nu_i \rightarrow e^-\bar\nu_i$ &  0.02 \\
$e^- e^+ \leftrightarrow \nu_i \bar \nu_i$ & 0.21 \\ \hline
$e^- \rightarrow \nu_i$  & 0.25 \\ \hline
$\nu_e\nu_i \rightarrow \nu_e\nu_i$ & 0.07 \\
$\nu_e\bar\nu_i \rightarrow \nu_e\bar\nu_i$ & 0.01 \\
$\nu_e \bar \nu_e \leftrightarrow \nu_i \bar \nu_i$ & 0.42 \\ \hline
$\nu_e \rightarrow \nu_i$ & 0.50 \\ \hline
\end{tabular}
\end{center}
\end{table}

In calculating the relative temperatures of $\nu_e$, $\nu_\mu$, and $\nu_\tau$
we must also consider the rate of energy transfer from the $\nu_e$ into the
other
two neutrino types, a process neglected in reference [6].  Again,
we can model these rates in terms of an energy loss term $w_{\nu_i}$,
$i = \mu$, $\tau$, where
\begin{equation}
\label{wvalue}
w_{\nu_i} = 2I_{\nu_e \rightarrow \nu_i}(T_{\nu_e} - T_{\nu_i})
\end{equation}
and $I_{\nu_e \rightarrow \nu_i}$ is a function of $T_{\nu_e}$ which
can be written in the form
\begin{equation}
\label{Ivalue2}
I_{\nu_e \rightarrow \nu_i} = \beta_{\nu_e \rightarrow \nu_i} G_F^2 T_{\nu_e}^8
\end{equation}
Again, there are three processes which contribute to the net
energy transfer from
$\nu_e$ into $\nu_i$:  $\nu_e \bar \nu_e \leftrightarrow \nu_i \bar \nu_i$,
$\nu_e \nu_i \rightarrow \nu_e \nu_i$, and $\nu_e \bar \nu_i \rightarrow
\nu_e \bar \nu_i$.  Using the standard matrix elements
(see, e.g., reference [8]), we have calculated the
differential cross sections for these three processes.  These cross sections
can then be substituted into the integral expressions for $I$ corresponding
to those given in reference [6] to derive the values for $\beta_{\nu_e
\rightarrow \nu_i}$ in
equation (\ref{Ivalue2}).  These values are given in Table 1.
Again, we see that $\nu_e \bar \nu_e$ annihilation is the dominant
energy transfer mechanism.
With this additional energy loss term, equation (\ref{rhonu}) becomes:
\begin{eqnarray}
\dot\rho_{\nu_e} &=& u_{\nu_e} -  \sum_{i=\nu,\tau} w_{\nu_i} -
4H\rho_{\nu_e},\\
\dot\rho_{\nu_i} &=& u_{\nu_i} +  w_{\nu_i} - 4H\rho_{\nu_i}~~~(i=\mu,\tau).
\end{eqnarray}

In deriving these equations, we have made two major approximations: first, we
have neglected the distortion in the energy density spectrum of the neutrinos,
treating the spectrum as a blackbody and considering only the change in the
total energy density.  This is relatively unimportant for the $\mu$ and $\tau$
neutrinos,
since their only effect on nucleosynthesis is via their contribution
to the total energy density, but it could be important for the electron
neutrinos, since the weak rates are highly energy dependent.  Second,
equations (\ref{uvalue}) and (\ref{wvalue}) are strictly valid only in the
limit
of $(T_\gamma - T_{\nu_i})/ T_\gamma \ll 1$ and $(T_{\nu_e} - T_{\nu_i}) /
T_{\nu_i} \ll 1$,
respectively; we have approximated the energy transfer rate by extrapolating
these equations into a regime where the temperature differences are large.
For instance, with $\tau = 1$ sec, and
for expansion times $t < 1$ sec
(the relevant time range for the freeze-out of the $n \leftrightarrow p$
reactions),
we find that
$(T_\gamma - T_{\nu_{\mu,\tau}}) / T_\gamma < 0.5$,
$(T_\gamma - T_{\nu_e}) / T_\gamma< 0.35$
and
$(T_{\nu_e} - T_{\nu_{\mu,\tau}}) / T_{\nu_e}   < 0.25$.
The temperature differences are smaller for shorter lifetimes.
For $\tau < 0.1$ sec we have
$(T_\gamma - T_{\nu_{\mu,\tau}}) / T_\gamma < 0.2$,
$(T_\gamma - T_{\nu_e}) / T_\gamma < 0.1$
and
$(T_{\nu_e} - T_{\nu_{\mu,\tau}}) / T_{\nu_e}   < 0.1$.
So we expect our approximation
to be quite accurate for $\tau < 0.1$ sec, but less accurate for larger
lifetimes.

We have integrated our temperature evolution equations for a decaying particle
with $rm_X = 10^4$ MeV.
Figure 1 shows the evolution of the neutrino temperatures as a function of time
for the
case of $\tau=0.1$ sec, including the effect of $e^-e^+$ annihilation.
The temperatures are displayed relative
to the photon temperature in terms of the fraction $T_{\nu_i}/T_\gamma$.
We see that for $t < \tau$, both $T_{\nu_e}/T_\gamma$ and
$T_{\nu_{\mu,\tau}}/T_\gamma$ decrease as the particle decays, but
the electron neutrinos are held closer
to the photon temperature than the $\mu$ and $\tau$ neutrinos, as in reference
[3].
The decrease in
$T_{\nu_i}/T_\gamma$ which occurs at $t > 10$ sec is the standard effect from
the annihilation of the $e^-e^+$ pairs and has nothing to do with the particle
decay.
An interesting ``rebound'' effect occurs between $10^{-1}$ and 1 sec.
The neutrino temperatures actually
increase briefly relative the photon temperature before the $e^-e^+$ entropy
dump sets in.  We believe that this is a real effect, not an artifact
of the approximations we have used.  It arises when the rate of energy
transfer to the photons (from the decaying particle) is greater than the
rate of energy transfer from the photons to the neutrinos.  Consider, for
example,
the (unphysical) limit where the energy of the decaying particles is
transferred instantaneously into the photons.  After a sharp increase
in the photon temperature, the ratio $T_\nu/T_\gamma$ would increase
as photon energy is transferred to the neutrinos.

Figure 2 shows the final temperature ratios, multiplied by
$(11/4)^{1/3}$ to factor out the effect of $e^-e^+$ annihilation,
as a function of X particle
lifetime.  Also shown for comparison are the results of the sudden decoupling
approximation used in reference [3].
This plot demonstrates significant differences between these two treatments.
The sudden decoupling approximation produces a large difference between the
electron and other neutrino temperatures in the region $\tau=0.1$ sec.  In our
more exact treatment, the differential heating effect is reduced, but it
persists
over a larger range in lifetime, for $\tau$ as large as 1 sec.
This is what one might have expected, since the more exact treatment allows
for energy transfer from photons to neutrinos at late times, when the
sudden decoupling approximation assumes that the neutrinos are already
decoupled.

\vskip 0.4 cm

\noindent {\bf III. EFFECTS ON $^4$HE PRODUCTION AND CONCLUSIONS}

We have applied these equations to the primordial nucleosynthesis
computer code of Wagoner [9], as modified by Kawano [10].  For a decaying
particle
with energy density $rm_X = 10^4$ MeV and lifetime $\tau$, we have calculated
the change in the primordial helium mass fraction $\Delta Y$ as a function
of $\tau$ for three values of the baryon to photon ratio:  $\eta = 10^{-10},
10^{-9.5},$ and $10^{-9}$.  These results are displayed in Fig. 3.  We find
that
$\Delta Y$ is nearly independent of $\eta$ and can be as large as $-0.012$.
We see that $\Delta Y < 0$ for $\tau < 1.5$ sec; for larger lifetimes the
decaying X particle, rather than its thermalized decay products, dominates
the expansion during the freeze-out of the $n \leftrightarrow p$ reactions,
resulting
in a larger expansion rate and a net increase in $^4$He.

For comparison, in Figure 4 we include the $\Delta Y$ value for $\eta =
10^{-9.5}$
using the sudden decoupling
approximation from reference [3].  We find that our more exact treatment
produces
a smaller decrease in the primordial helium, but that the effect occurs
over a larger range in particle lifetime; we find that $\Delta Y < -0.005$
for $\tau = 0.05 - 1.2$ sec.  This is not surprising, since
our more exact treatment produces a smaller differential heating effect spread
out over a larger range in $\tau$.  Part of the difference, however, is that
the actual total interaction rates used to
compute decoupling in references [3]-[4] are different from those
used here.  So in Figure 4 we also
give the helium abundances corresponding to sudden decoupling
when we change the total interaction rates in
equation (\ref{Gammaold}) to give the same freeze-out temperatures as those
derived in ref. [6]
($\alpha_{\nu_e} = 1.5$, $\alpha_{\nu_{\mu,\tau}} = 0.33$).
The curve is shifted over to correspond more closely to our more exact
treatment, but it retains a much sharper and deeper minimum.

Of the $\nu_\tau$ decay modes considered by Dodelson et al. [2],
our present work most closely
resembles $\nu_\tau$ decay into sterile plus electromagnetic decay products.
No reduction
in $^4$He is evident for this case in reference [2], but this is
to be expected, because the $\nu_\tau$ energy density is too low
to lead to significant differential neutrino heating.  Dodelson et al. [2]
do see a significant reduction in $^4$He for decay modes which include
a $\nu_e$ in the final state, but this reduction occurs for entirely
different reasons:  the additional $\nu_e$ decay products increase
the $n \leftrightarrow p$ weak rates, keeping them in thermal equilibrium
longer.

Our results indicate that the decrease in $^4$He production due to differential
neutrino heating from a decaying particle, first discussed in reference [3],
holds up under a more exact treatment.  The effect is not as large as had
been previously estimated, but it occurs over a longer range in particle
lifetime.  The change in $Y$ can be as large as $-0.01$ for $\tau$ in the
range $0.1  - 0.7$ sec, and as large as $-0.005$ for $\tau = 0.05 - 1.2$ sec.
A decrease on the order of 0.01 can resolve
current problems with standard primordial nucleosynthesis [5], although
it is perhaps not the most plausible mechanism for resolving these problems.
A more detailed treatment of the spectral distortion using the
full machinery of reference [2] seems justified, although we do
not expect our results to be significantly altered in such a treatment.
Even the treatment in reference [2] would need to be modified to treat this
problem correctly, since that treatment also assumes small temperature
differences.

\vskip 0.4 cm

\noindent {\bf ACKNOWLEDGMENTS}

We thank S. Dodelson for helpful discussions.
R.E.L. and R.J.S.
were supported in part by the Department of
Energy (DE-AC02-76ER01545).  R.J.S. was supported in part by NASA (NAG 5-2864).

\vfill
\eject

\centerline {\bf References}

\vskip 0.5 cm
\begin{description}
\item {[1]}M. Kawasaki, P. Kernan, H.-S. Kang, R.J. Scherrer, G. Steigman,
and T.P. Walker, Nucl. Phys. B {\bf 419}, 105 (1994).

\item {[2]}S. Dodelson, G. Gyuk, and M.S. Turner, Phys. Rev. D {\bf 49},
5068 (1994).

\item {[3]}R.J. Scherrer and M.S. Turner, Ap.J., {\bf 331}, 19 (1988).

\item {[4]}R.J. Scherrer, J. Cline, S. Raby, \& D. Seckel, Phys. Rev. D, {\bf
44},
3760 (1991).

\item {[5]}C. Copi, D.N. Schramm, and M.S. Turner, Science, {\bf 267}, 192
(1995);
N. Hata, R.J. Scherrer, D. Thomas, T.P. Walker, S. Bludman, and P. Langacker,
Phys. Rev. Lett., submitted; C. Copi, D.N. Schramm, and M.S. Turner, Phys. Rev.
Lett.,
submitted.

\item {[6]}N.C. Rana and B. Mitra, Phys. Rev. D, {\bf 44}, 393 (1991).

\item {[7]}A.M. Herrera and S. Hacyan, Phys. Fluids {\bf 28}, 3253 (1985);
A.M. Herrera and S. Hacyan, Astrophys. J. {\bf 336}, 539 (1989).

\item {[8]}S. Dodelson and M.S. Turner, Phys. Rev. D, {\bf 46}, 3372 (1992).

\item {[9]}R.V. Wagoner, W.A. Fowler, and F. Hoyle, Astrophys. J. {\bf 148},
3 (1967); R.V. Wagoner, Astrophys. J. Suppl. Ser. {\bf 162}, 247 (1969);
Astrophys. J. {\bf 179}, 343 (1973).

\item {[10]} L. Kawano, FERMILAB-Pub-92/04-A (1992).

\end{description}

\vfill
\eject

\centerline{\bf FIGURE CAPTIONS}

\vskip 0.5 cm

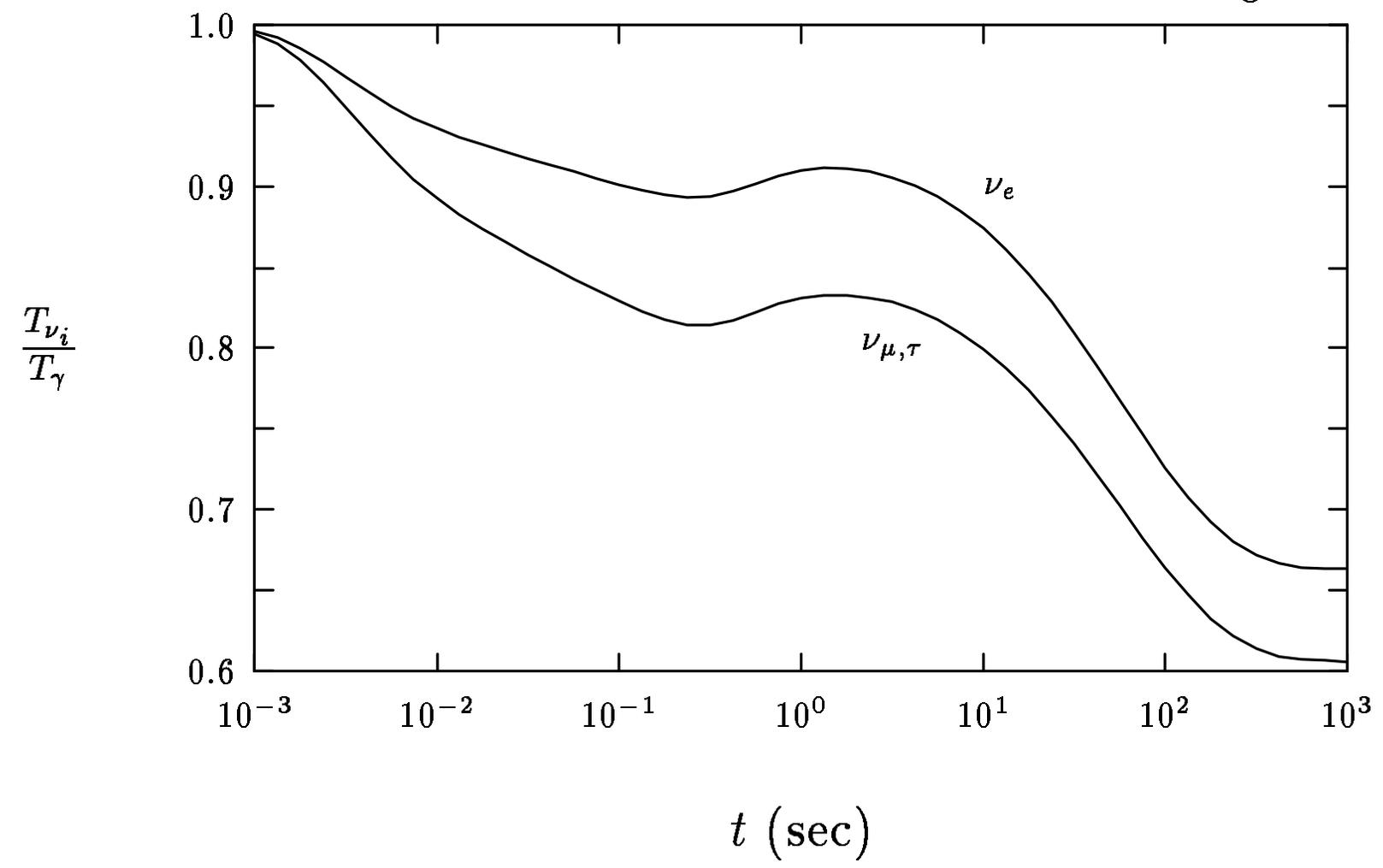
\noindent FIG. 1.  The time evolution of $T_{\nu_i}/T_\gamma$ for
electron neutrinos (upper curve) and $\mu$ and $\tau$ neutrinos (lower curve)
with a decaying particle with lifetime $\tau = 0.1$ sec and energy
density parameter $rm_X = 10^4$ MeV.

\vskip 0.5 cm

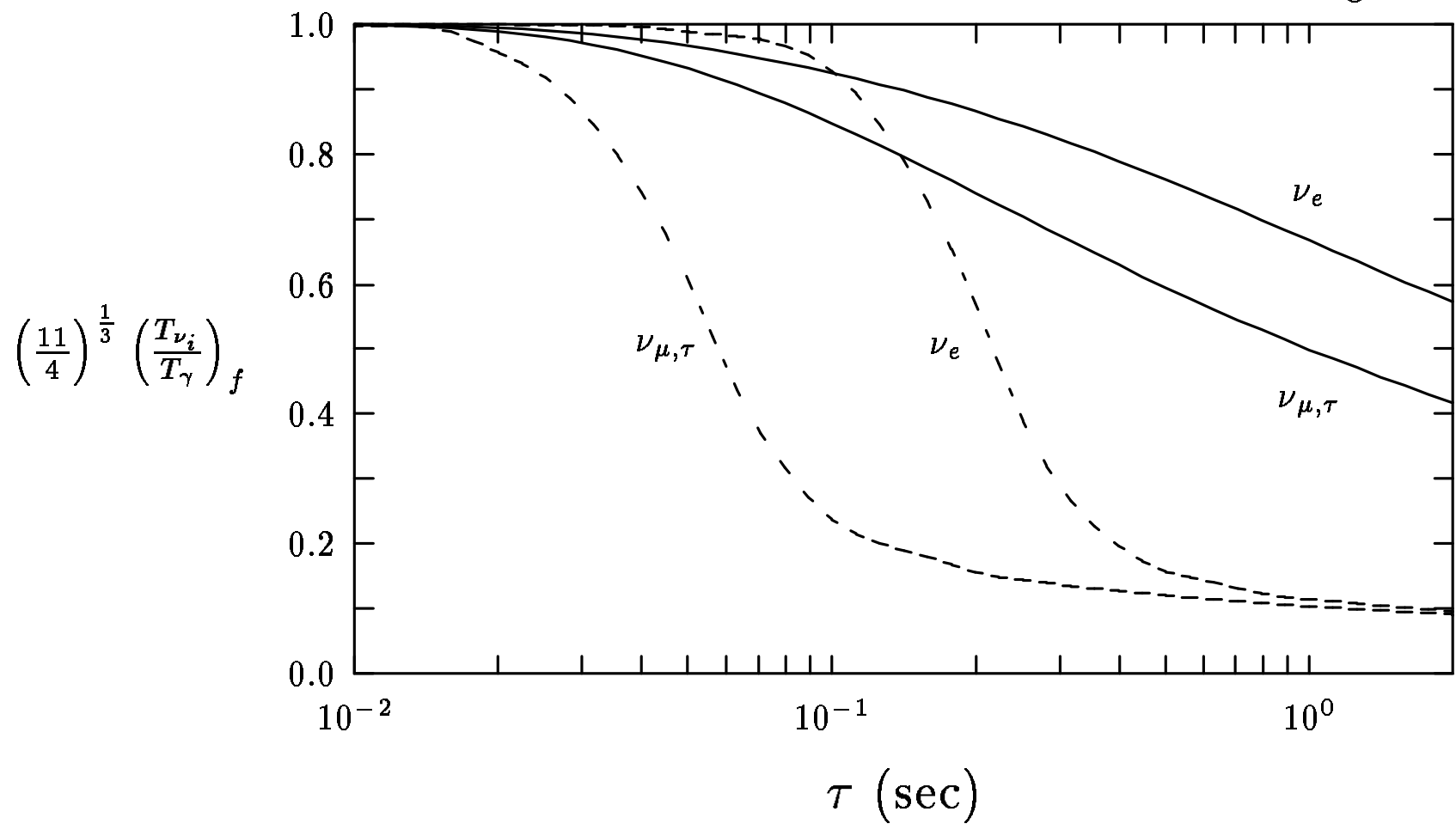
\noindent FIG. 2. The final ratios $(11/4)^{1/3} (T_{\nu_i}/T_\gamma)$
as a function of particle lifetime $\tau$ for
a decaying particle with a density parameter $rm_X = 10^4$ MeV (solid curves).
For comparison, the corresponding ratios are also shown for the sudden
decoupling approximation [3] (dashed curves).
\vskip 0.5 cm

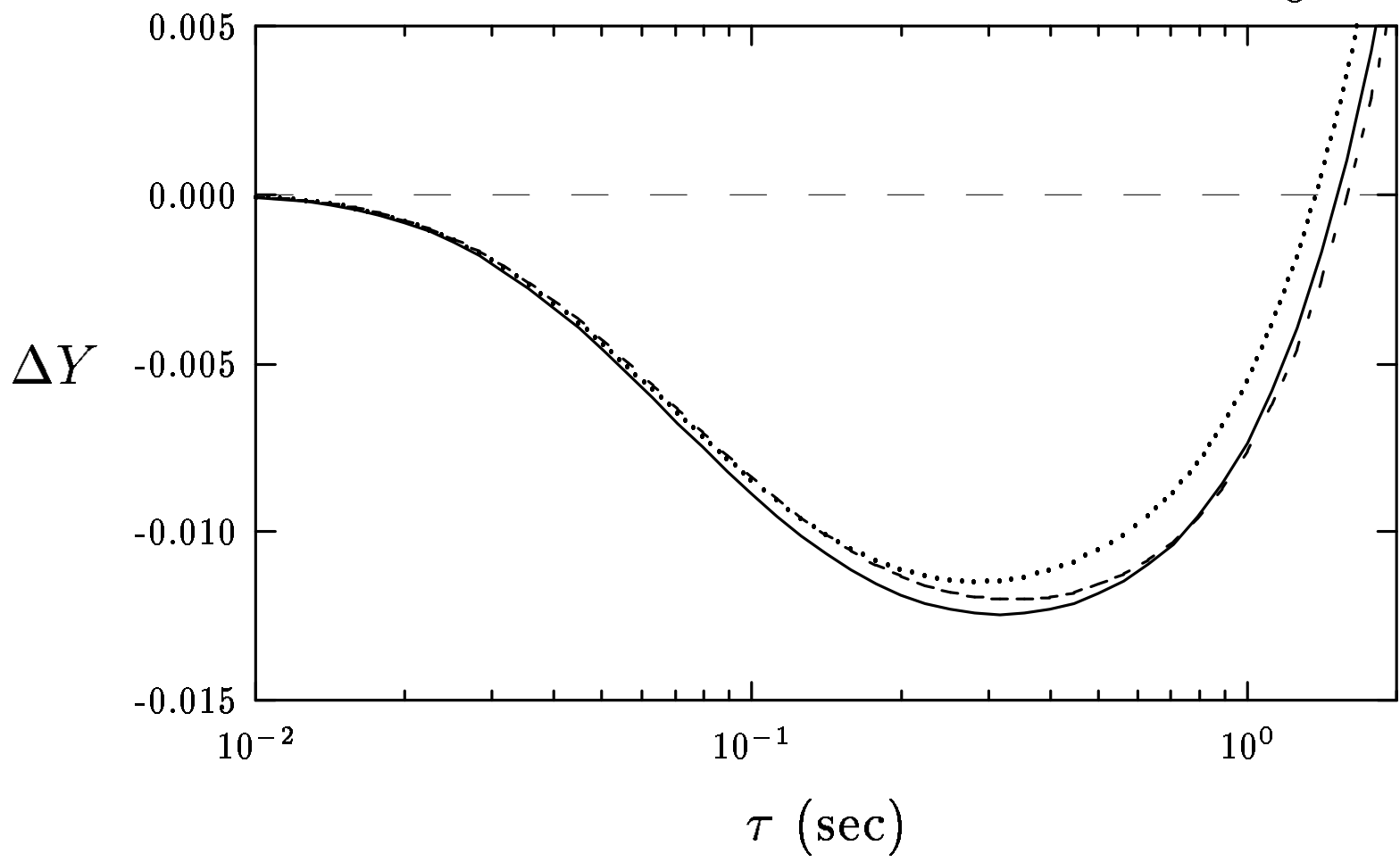
\noindent FIG. 3. The change in the primordial $^4$He abundance, $\Delta Y$,
as a function of decaying particle lifetime $\tau$
for
a decaying particle with a density parameter $rm_X = 10^4$ MeV
and
three different baryon-to-photon ratios:  $\eta = 10^{-10}$ (dashed curve),
$\eta = 10^{-9.5}$ (solid curve) and $\eta = 10^{-9}$ (dotted curve).

\vskip 0.5 cm

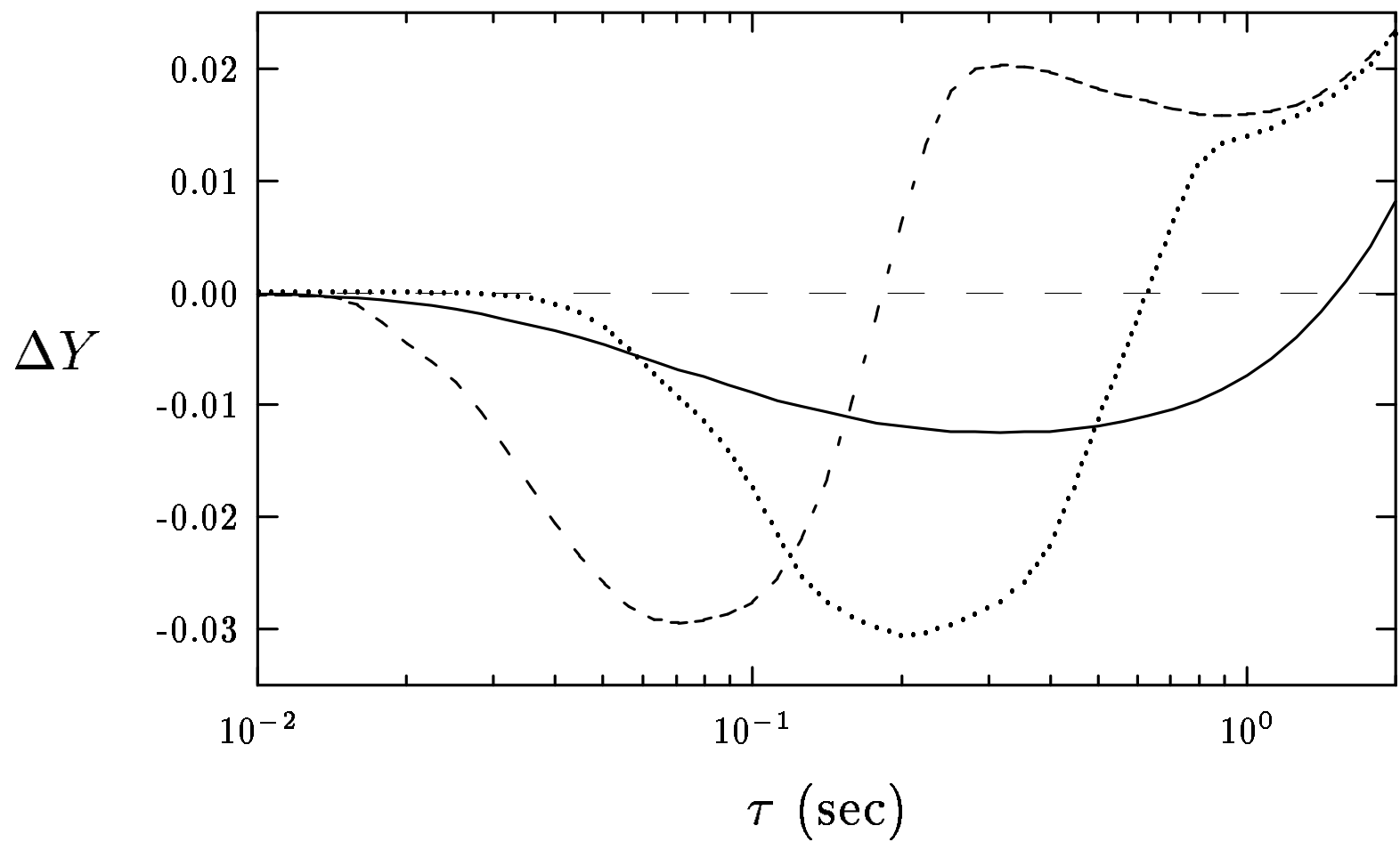
\noindent FIG. 4.  The change in the primordial $^4$He abundance, $\Delta Y$,
for $\eta = 10^{-9.5}$,
as a function of decaying particle lifetime $\tau$
for
a decaying particle with a density parameter $rm_X = 10^4$ MeV
(solid curve) compared with the results of the sudden decoupling approximation
[3] (dashed curve) and the sudden decoupling approximation with the
interaction rates updated to correspond to the ones used here (dotted curve).

\end{document}